\documentclass{elsart}

\usepackage{cite}

\begin{document}

\begin{frontmatter}

\title{Comment on: ``On the Klein-Gordon oscillator subject to a Coulomb-type
potential''. Ann. Phys. 355 (2015) 48}

\author{Francisco M. Fern\'{a}ndez}

\address{INIFTA, Divisi\'{o}n Qu\'{i}mica Te\'{o}rica,\\
Blvd. 113 y 64 (S/N), Sucursal 4, Casilla de Correo 16,\\
1900 La Plata, Argentina} \thanks{ E-mail:
fernande@quimica.unlp.edu.ar}

\begin{abstract}
We analyze the conclusions of the influence of a Coulomb-type potential on
the Klein-Gordon oscillator. We show that the truncation method proposed by
the authors do not yield all the eigenvalues of the radial equation but just
one of them for a particular value of a model parameter. Besides, the
existence of allowed oscillator frequencies that depend on the quantum
numbers is an artifact of the truncation method.
\end{abstract}


\end{frontmatter}

In a paper published in this journal Bakke and Furtado\cite{BF15} study the
influence of a Coulomb-type potential on the Klein-Gordon oscillator. By a
series of transformations of the Klein-Gordon equation in cylindrical
coordinates these authors derive an eigenvalue equation for the radial part
of the wavefunction. By means of the Frobenius method they obtain a
three-term recurrence relation for the coefficients of the power series. A
suitable truncation of the series enables them to derive an analytical
expression for the energy levels and a most striking result: the existence
of allowed oscillator frequencies. The purpose of this Comment is the
analysis of the truncation method used by Bakke and Furtado and its effect
on the conclusions drawn in their paper.

The starting point is the Klein-Gordon equation
\begin{equation}
-\frac{\partial \phi }{\partial t^{2}}+\frac{\partial ^{2}\phi }{\partial
\rho ^{2}}+\frac{1}{\rho }\frac{\partial \phi }{\partial \rho }+\frac{1}{%
\rho ^{2}}\frac{\partial ^{2}\phi }{\partial \varphi ^{2}}+m\omega \phi
-m^{2}\omega ^{2}\rho ^{2}\phi -m^{2}\phi -\frac{2mf}{\rho }\phi -\frac{f^{2}%
}{\rho ^{2}}\phi =0,  \label{eq:Schro_fime_dep}
\end{equation}
where the meaning of each parameter is given in the authors' paper\cite{BF15}%
. Units are chosen so that $c=\hbar =1$ (in a recent paper\cite{F20} we have
critiziced this unclear choice of suitable units). The authors look for a
particular solution of the form
\begin{equation}
\phi (t,\rho ,\varphi )=e^{-i\mathcal{E}t}e^{il\varphi }R(\rho ),
\label{eq:particular solution}
\end{equation}
where $l=0,\pm 1,\pm 2,\ldots $. In this way they obtain the following
equation for the radial function $R(\rho )$
\begin{equation}
\frac{d^{2}R}{d\rho ^{2}}+\frac{1}{\rho }\frac{dR}{d\rho }-\frac{\gamma ^{2}%
}{\rho ^{2}}R-\frac{2mf}{\rho }R-m^{2}\omega ^{2}\rho ^{2}R+\beta ^{2}R=0,
\label{eq:radial_eq_1}
\end{equation}
where $\beta ^{2}=\mathcal{E}^{2}-m^{2}+m\omega $ and $\gamma
^{2}=l^{2}+f^{2}$.

By means of the change of variables $\xi =\sqrt{m\omega }\rho $ one obtains
\begin{equation}
\frac{d^{2}R}{d\xi ^{2}}+\frac{1}{\xi }\frac{dR}{d\xi }-\frac{\gamma ^{2}}{%
\xi ^{2}}R-\frac{\delta }{\xi }R-\xi ^{2}R+\frac{\beta ^{2}}{m\omega }R=0,
\label{eq:radial_eq_2}
\end{equation}
where $\delta =\frac{2mf}{\sqrt{m\omega }}$. If we write the solution to
this equation as
\begin{equation}
R(\xi )=\xi ^{|\gamma |}e^{-\frac{\xi ^{2}}{2}}\sum_{j=0}^{\infty }a_{j}\xi
^{j},  \label{eq:Frobenius}
\end{equation}
we obtain the three-term recurrence relation
\begin{eqnarray}
a_{j+2} &=&\frac{\delta }{(j+2)(j+1+\alpha )}a_{j+1}-\frac{\theta -2j}{%
(j+2)(j+1+\alpha )}a_{j},  \nonumber \\
j &=&-1,0,1,2\ldots ,\;a_{-1}=0,\;a_{0}=1,  \label{eq:rec_rel}
\end{eqnarray}

The authors state that ``It is well-known that the quantum theory requires
that the wave function (6) must be normalizable. Therefore, we assume that
the function $R(\xi )$ vanishes at $\xi \rightarrow 0$ and $\xi \rightarrow
\infty $. This means that we have a finite wave function everywhere, that
is, there is no divergence of the wave function at $\xi \rightarrow 0$ and $%
\xi \rightarrow \infty $, then, bound state solutions can be obtained.
However, we have written the function $H(\xi )$ as a power series expansion
around the origin in Eq. (14). Thereby, bound state solutions can be
achieved by imposing that the power series expansion (14) or the Heun
biconfluent series becomes a polynomial of degree $n$. In this way, we
guarantee that $R(\xi )$ behaves as $\xi ^{|\gamma |}$ at the origin and
vanishes at $\xi \rightarrow \infty $. Through the recurrence relation (15),
we can see that the power series expansion (14) becomes a polynomial of
degree $n$ by imposing two conditions:
\begin{equation}
g=2n\ \mathrm{and}\ a_{n+1}=0,  \label{eq:truncation_cond}
\end{equation}
where $n=1,2,\ldots $.''

It clearly follows from these two conditions that $a_{j}=0$ for all $j>n$.
However, the authors' statement is a gross conceptual error because a bound
state simply requires that $R(\xi )$ is square integrable:
\begin{equation}
\int_{0}^{\infty }\left| R(\xi )\right| ^{2}\xi \,d\xi <\infty .
\label{eq:bound-state_def_xi}
\end{equation}
Consequently, by means of the truncation condition (\ref{eq:truncation_cond}%
) the authors obtain just some particular bound states and only for
particular values of a model parameter. For example, when $n=1$ then $\theta
=2$ and we obtain a simple analytical expression for $\mathcal{E}_{1,l}$.
The second condition $a_{2}=0$ yields $\delta =\delta _{1,l}=\pm \sqrt{%
2\alpha }$ and the oscillator frequency $\omega _{1,l}=4mf^{2}/\delta
_{1,l}^{2}$.

From the general result the authors argue as follows ``... we have
that the influence of the Coulomb-like potential makes that the
ground state to be defined by the quantum number $n=1$ instead of
the quantum number $n=0$. Note that we have written the angular
frequency $\omega $ in terms of the quantum numbers $\{n,l\}$ in
Eq. (18). From the mathematical point of view, this dependence of
the angular frequency of this relativistic oscillator on the
quantum numbers $\{n,l\}$ results from the fact that the exact
solutions to Eq. (12) are achieved for some values of the
Klein-Gordon oscillator frequency. From the quantum mechanics
point of view, this is an effect which arises from the influence
of the Coulomb-type potential on the Klein-Gordon oscillator.''
And also ``The meaning of achieving this relation is that not all
values of the angular frequency $\omega $ are allowed, but some
specific values of $\omega $ which depend on the quantum numbers
$\{n,l\}$. For this reason, we label $\omega =\omega _{n,l}$.''

In order to put the authors' conclusions to the test, we solve the
eigenvalue equation (\ref{eq:radial_eq_2}) for $W=\frac{\beta ^{2}}{m\omega }
$ so that
\begin{eqnarray}
\mathcal{E}^{2} &=&m\omega W+m^{2}-m\omega ,  \nonumber \\
W &=&\theta +2\left( |\gamma |+1\right) .  \label{eq:E^2}
\end{eqnarray}
Anybody familiar with eigenvalue equations realizes that for a given set of
values of $f$, $\delta $ and $l$ we obtain an infinite set of eigenvalues $%
W_{\nu ,l}(f,\delta )$, $\nu =0,1,\ldots $, for those solutions $R_{\nu
,l}(\xi )$ that satisfy equation (\ref{eq:bound-state_def_xi}). A little
thinking reveals that the dependence of the angular frequency on the quantum
numbers is an artifact of the truncation of the power series by means of the
condition (\ref{eq:truncation_cond}) and it is expected to occur when
looking for exact solutions to conditionally solvable problems\cite{D88,
BCD17} (and references therein). If one solves the eigenvalue equation (\ref
{eq:radial_eq_2}) in a proper way such dependence does not take place.

Since the eigenvalue equation (\ref{eq:radial_eq_2}) is not
exactly solvable (contrary to what the authors appear to believe)
we resort to the reliable Rayleigh-Ritz variational method that is
well known to yield increasingly
accurate upper bounds to all the eigenvalues of the Schr\"{o}dinger equation%
\cite{P68} (and references therein). For simplicity, we choose the basis set
of (unnormalized) functions $\left\{ u_{j}(\xi )=\xi ^{|\gamma |+j}e^{-\frac{%
\xi ^{2}}{2}},\;j=0,1,\ldots \right\} $. In order to test the accuracy of
the variational results we also apply the powerful Riccati-Pad\'{e} method%
\cite{FMT89a}. As a suitable particular case we arbitrarily choose $l=0$ and
$f^{2}=1$ so that $\gamma ^{2}=1$ and $\alpha =3$. From the truncation
condition (\ref{eq:truncation_cond}) with $n=1$ we obtain $\theta =2$, $%
W_{BF}=6$ and $\delta =\pm \sqrt{6}$.

When $\delta =\sqrt{6}$ ($f=1$) the two methods mentioned above yield $%
W_{0,0}=6$, $W_{1,0}=9.805784090$, and $W_{2,0}=13.66928892$ for the three
lowest eigenvalues. It is clear that the truncation method only yields the
lowest eigenvalue and misses all the other ones. On the other hand, when $%
\delta =-\sqrt{6}$ ($f=-1$) the two methods yield $W_{0,0}=1.600357154$, $%
W_{1,0}=6$, $W_{2,0}=10.21072810$. In this case the truncation method only
gives us the second lowest eigenvalue and misses all the other ones. It is
clear that the value of $n$ in the truncation condition (\ref
{eq:truncation_cond}) is not related with the quantum number $\nu $ used to
arrange the eigenvalues in increasing order of magnitude.

The authors' equation for the energy levels suggest that $\mathcal{E}%
_{n,l}(-f)=$ $\mathcal{E}_{n,l}(f)$; however, such reflection symmetry only
applies to the particular energy levels stemming from the truncation
condition (\ref{eq:truncation_cond}). Present numerical calculations show
that in general $W_{\nu ,l}(-f)\neq W_{\nu ,l}(f)$.

Summarizing: the truncation condition (\ref{eq:truncation_cond}) only
provides one energy level $W_{n,l}$ (or $\mathcal{E}_{n,l}^{2}$) for a
particular value of of $\delta =\delta _{n,l}$ and misses all the other
energy levels $W_{\nu ,l}$ for that value of $\delta $. Besides, it fails to
provide the energy levels for other values of that parameter. Consequently,
the dependence of the oscillator frequency on the quantum numbers is a mere
artifact of the truncation condition (\ref{eq:truncation_cond}). This
artificial condition is not necessary for the existence of bound states that
should satisfy the well-known (and widely more general) condition (\ref
{eq:bound-state_def_xi})\cite{CDL77}.

\end{document}